%% file: turb_front.tex
\documentclass[prl,a4paper,twocolumn,showpacs,superscriptaddress,amssymb]{revtex4}
\usepackage{bm}
\usepackage{graphicx}

\usepackage[dvips]{color} 

\begin{document}
\input mydefs.tex

\title{Quantum turbulence in propagating superfluid vortex front }

\author{V.B.~Eltsov}
\affiliation{Low Temperature Laboratory, Helsinki University of Technology, P.O.Box
2200, 02015 HUT, Finland }
\affiliation{Kapitza Institute for Physical Problems,
Kosygina 2, 119334 Moscow,  Russia}

\author{A.I.~Golov}
\affiliation{Department of Physics and Astronomy, Manchester University,
  Manchester, United Kingdom}

\author{R.~de~Graaf}
\affiliation{Low Temperature Laboratory, Helsinki University of Technology,
 P.O.Box 2200, 02015 HUT, Finland }

\author{R.~H\"anninen}
\affiliation{Low Temperature Laboratory, Helsinki University of
Technology, P.O.Box 2200, 02015 HUT, Finland }

\author{M.~Krusius}
\affiliation{Low Temperature Laboratory, Helsinki University of Technology, P.O.Box
2200, 02015 HUT, Finland }

\author{V.S.~L'vov}
\affiliation{Department of Chemical Physics, The Weizmann Institute of
  Science, Rehovot 76100, Israel}

\author{R.E.~Solntsev}
\affiliation{Low Temperature Laboratory, Helsinki University of Technology, P.O.Box
2200, 02015 HUT, Finland }

\date{\today}

\begin{abstract}
We present experimental, numerical and theoretical studies of a
vortex front propagating into a region of vortex-free flow of rotating
superfluid $^3$He-B.  We show that the nature of the front changes
from laminar through   quasi-classical turbulent to quantum
turbulent with decreasing temperature. Our experiment provides the
first direct measurement of the dissipation rate in turbulent
vortex dynamics of $^3$He-B and demonstrates that the dissipation
is temperature- and mutual friction-independent in the
$T\rightarrow 0$ limit, and is strongly suppressed when the
Kelvin-wave cascade on vortex lines is predicted to be involved in
the  turbulent energy transfer to smaller length scales.
\end{abstract}
\pacs{67.57.Fg, 47.32.-y, 67.40.Vs} \maketitle 
Turbulent motion of fluids with low viscosity, like water or air
is    a general phenomenon in Nature and   plays an important role
in human everyday life. Nevertheless, various features of
turbulence are not yet well understood; thus a possibility to
study turbulence from another, non-classical viewpoint looks
promising. Flows in superfluids (with zero viscosity) can be
turbulent; understanding this form of turbulence  is crucial for
our ability to describe superfluids,   including     practical
applications like cooling superconductor devices.  Turbulence in
superfluid $^4$He at relatively high  temperatures $T\sim 1\,$K has
been studied for decades~\cite{VinenNiemela}. Fermi-fluid $^3$He
is very different from the Bose-fluid $^4$He; this allows us to
study in $^3$He some aspects of superfluid turbulence not
available in $^4$He \cite{exp4,dyn-review}.

\begin{figure}[top]
\centerline{\includegraphics[width=\linewidth]{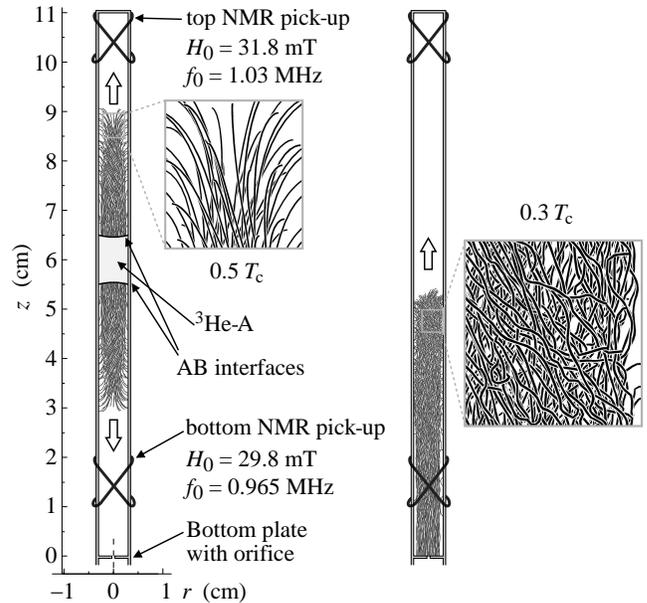}}
  \caption{\label{f:setup} Experimental setup. \textit{(Left)}
  Configuration for vortex injection using AB interface
  instability in which  two vortex fronts
  propagate  independently in the upper and lower B phase sections (as shown by hollow arrows).
  \textit{(Right)} Configuration for vortex injection from the
  orifice when a single vortex front travels first through the
  bottom and then through the top NMR pick-up coils. Superimposed
  inside the sample container are snapshots of vortex
  configurations from numerical simlulations of vortex expansion
  at two different temperatures. The configuration at $0.3T_{\rm
  c}$ displays small-scale structure (shown within blow-up on the right) which is
  absent at $0.5T_{\rm c}$. }
\end{figure}

In this Letter we report the first experimental observation and
study of propagating   laminar and turbulent vortex fronts in
rotating $^3$He which opens a possibility to directly measure
the rate of kinetic energy dissipation in a wide temperature range
down to $0.18\,T\sb c$ of the critical temperature $T\sb c\approx
2.43\,  $mK (at $29$ bar pressure). Theoretical and numerical
analysis of the propagating   front   allows us to identify the
mechanisms of   energy dissipation and to discover a
dissipation anomaly: the saturation of the rate of turbulent
kinetic energy dissipation at a nonzero value at temperatures
below $~0.22\, T\sb c$.

The idea of the experiment is as follows: in rotating $^3\!$He (in the
B-phase) we can create a vortex-free Landau state, which in the absence of
external perturbations persists forever.  In this state the superfluid
component is at rest in the laboratory frame, while the normal component is
in rigid rotation with angular velocity $\O$. The Landau state is
metastable, having larger free energy than the stable equilibrium vortex state.
The latter consists of rigidly rotating normal and superfluid components
with a regular array of rectilinear quantized vortices. When we inject a
seed vortex into the Landau state, we observe a rapid local evolution of
the vorticity toward the equilibrium state. A boundary between the
vortex-free and the vortex states propagates with constant velocity $V\sb
f$ toward the metastable region(s), see~\Fig{f:setup}. In some sense this
phenomenon is similar to the propagation of a flame front in premixed fuel.
The flame front propagation can also proceed in laminar or in turbulent
regimes. In the latter case the effective area of the front increases and
its propagation speed becomes higher than in the laminar regime. This
property finds its practical use in combustion engines but also has been
used to describe intensity curves of type Ia supernovae~\cite{SN}. In all
these cases a metastable state of matter is converted to the stable state in the
front and $V\sb f$ is determined by the rate of dissipation of the released
energy. In our case (with density $\r\sb s$ and
velocity $U\sb s=\O\, r$ of the superfluid component in a sample container of radius $R$) the
dissipation rate of total kinetic energy, $\C E(t)$, is related to $V\sb f$
as 
\BE{2} {d \C E}\big/ {d t} = - \pi \r\sb s V\sb f\, \O^2 R^4/4\ .  \ee 
Thus by measuring $V\sb f$ one determines directly ${d \C E} / {d t}$ in
turbulent vortex dynamics. So far measurements in the turbulent regime only
concerned the vortex density~\cite{exp4}.

Our main experimental   observation is that ${d \C E(t)}\big/ {d
t}$ does not go to zero in the $T\rightarrow0$ limit. At high
enough $T$ the front is laminar and its velocity is determined by
mutual friction between the normal and superfluid components
$V_{\rm f}(T) \approx \alpha(T)\, \Omega R  $, where   $\a(T)$ is
a dimensionless mutual-friction coefficient. Although $\alpha(T)\rightarrow 0$ when $T\rightarrow0$,
the measured velocity $V_{\rm f}(T)$ saturates at a constant value which
corresponds to an effective friction $\alpha_{\rm eff} \sim 0.1$.
We interpret this behavior to be similar to the {\it viscous
anomaly} in classical turbulence, where the dissipation rate does
not vanish when viscosity $\nu$ goes to zero.   The viscous
anomaly appears due to cascading of energy to smaller length
scales. When $\nu\rightarrow 0$ the smallest length scale
decreases but the global dissipation rate does not change. We
believe that a similar mechanism, which we can call {\it mutual
friction anomaly}, applies to superfluids    where mutual friction
plays the role of viscosity. One can say that the viscous and
mutual friction anomalies are particular cases of a more general
phenomenon, which we call the {\it dissipation anomaly}: The
nonzero rate of energy dissipation in the limit of a vanishingly
small parameter, that governs dissipation.

As mutual friction decreases and turbulent motion reaches
progressively smaller length scales, eventually quantized vortex
lines become important. The energy cascade on   length scales
smaller than the intervortex distance  and the nature of
dissipation  on such scales are currently the central questions in
research on turbulence in superfluids~\cite{VinenNiemela}.  At the
moment only theoretical speculations exist on the role of the
non-linear interaction of Kelvin waves, resulting in a Kelvin-wave
cascade~\cite{svistunov,Vin03-PRL},   terminated by quasiparticle
emission, and on the role of vortex reconnections which can
immediately redistribute energy over a range of scales and also
lead to dissipation~\cite{5}. Our experiments show evidence for
the importance of the Kelvin-wave cascade:  We   have observed a
rapid decrease of the front velocity with decreasing temperature
in the region where the introduction of sub-intervortex scales to
energy transfer is expected. We propose an explanation of this
effect using a model with a bottleneck crossover between
quasi-classical (at  super-intervortex scales) and quantum
turbulence  (at sub-intervortex scales)~\cite{LNR}.

\begin{figure}
\centerline{\includegraphics[width=0.9\linewidth]{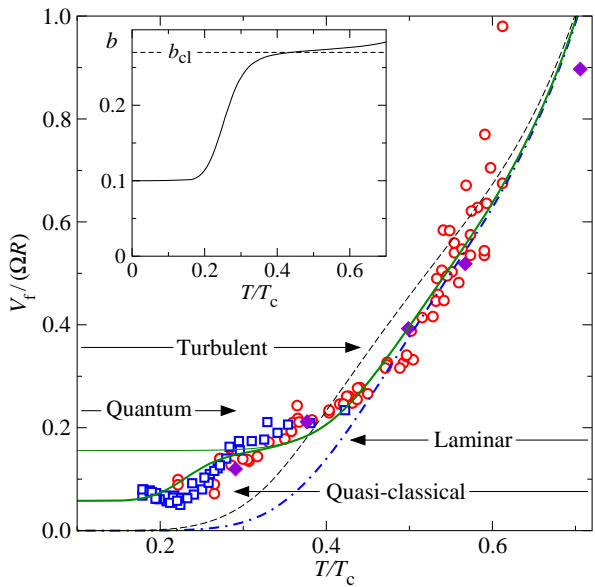}}
\caption{Color online. Scaled front velocity vs
temperature. Open circles and squares mark measurements with
injection using AB interface instability and orifice trapping,
respectively, at $\O$ between 0.8 and 1.5\,rad/s.
 Dashed line is the measured value of mutual friction coefficient
 $\alpha(T)$~\cite{bevan-friction}, extrapolated below $0.35\,T_{\rm c}$
 with $\exp(-\Delta/T)$ law~\cite{delta}. Filled diamonds are results of
numerical simulations. Dash-dotted, thin solid and thick solid
lines show model approximations which sequentially account for
dissipation in the large-scale motion, turbulent energy transfer
and bottleneck effect. \textit{(Insert)} Value of parameter $b(T)$
in Eq.~(\ref{5}) which was used in producing the thick solid line
in the  main panel.
  \label{f:Vf}}
\end{figure}

\textit{Experiment:} Our measurements are performed in a
rotating nuclear demagnetization cryostat at $\O\sim 1 $rad/s.  The
$^3$He-B sample at 29\,bar pressure is contained in a cylindrical cell with
radius $R=3\,$mm and length 110\,mm, oriented parallel to the rotation
axis, Fig.~\ref{f:setup}.  Pick-up coils of two independent NMR
spectrometers near the top and bottom of the cell are used to monitor the
vortex configuration~\cite{dyn-review}. To prepare the initial vortex-free
state we heat the sample to about 0.75\,$T_{\rm c}$ (for rapid annihilation
of all vortices) and then cool it in the vortex-free state in rotation to
the target temperature. We can inject seed vortices in the middle of the
cell, using the instability of the AB interface in rotation
\cite{dyn-review}, controlled with an applied magnetic field. In this case
two vortex fronts propagate independently up and down, arriving to the top
and bottom pick-up coils practically simultaneously. A second injection
technique uses remanent vortices which are trapped in the vicinity of the
orifice on the bottom of the sample, Fig.~1~right. In this case the vortex
front propagates upwards along the entire sample through both pick-up coils
in succession.  We determine the front velocity dividing the flight
distance by the flight time, as if the front propagates in
steady-state configuration. Although this is not the case due
to initial equilibration processes which follow injection, we believe
that this simplification is justified here, since the two
injection techniques for different propagation lengths give the same
result, as seen in Fig.~\ref{f:Vf}.

\textit{Experimental results} on the front velocity  are
presented in Fig.~\ref{f:Vf}. The temperature range is clearly
divided in two regions: At $T\gtrsim 0.4 T_{\rm c}$ the
dimensionless front velocity ${\cal V}_{\rm f} = V_{\rm f}/\Omega
R \approx \alpha$. This agrees with previous measurements in this
temperature range and can be understood from the dynamics of a
single vortex, when inter-vortex interactions are
ignored \cite{old-flight-meas}. We call this region the laminar
regime. The measured values of ${\cal V}_{\rm f}$ are slightly
below $\alpha$.  We believe that the difference is caused by the
twisted vortex state~\cite{twisted} behind the front. It reduces
the energy difference across the front and correspondingly the
front velocity. Estimating the reduction factor from the uniform
twist model \cite{twisted} we get ${\cal V}_{\rm f,lam} =
[2/\log(1+1/q^2) - 2/q^2]\alpha$, where $q = \alpha/(1-\alpha')$
and $\alpha'$ is the reactive mutual friction coefficient. This
dependence, (the dash-dotted line in Fig.~\ref{f:Vf})  is in good
agreement with the experiment.

The new feature is the behavior at $T<0.4\, T_{\rm c}$. Here
$V_{\rm f}$ rapidly deviates to  larger values than $\alpha \Omega
R$ and eventually becomes  constant  in the $T\rightarrow 0$ limit
with a peculiar transition from one plateau to another at around
$0.25\,T_{\rm c}$. We attribute this behavior to turbulent
dynamics and analyze it below in more detail.

\textit{Simulations:} To  clarify    the vortex front
formation and propagation,   we simulate  vortex dynamics using the
vortex filament model with full Bio-Savart equations and an
additional solution of the Laplace   equation for solid wall
boundary conditions~\cite{simul}  in an ideal  cylinder with
length 40\,mm and diameter 3\,mm rotating at $\O=1$rad/s. In the
initial configuration the equilibrium number  of vortices is
placed close to one end plate of the sample as quater-loops between
the end plate and the side wall of the sample.  During evolution
vortices  form a front propagating along the sample (for movies
see Ref.~\cite{movie}). The thickness of the front is
$\Delta(r)\simeq r \,d$, where $d\sim 1$. For $T>0.4 T\sb c$  the
front is laminar with smooth  vortices, that  only twist at large
scales (Fig.~\ref{f:setup}, left). For smaller $T$ the front
becomes turbulent demonstrating a lot of small-scale vortex
structure: Kelvin waves, kinks, semi-loops (Fig.~\ref{f:setup},
right). These small scale fluctuations exist  on the background of
strongly polarized vortex orientation, which is still preserved at
low temperatures. The calculated front velocity is in good
agreement with the measurements, see Fig.~\ref{f:Vf}.  In
particular, the velocity is   lower than $\alpha \, \Omega R$ at
higher temperatures and propagation becomes relatively much faster
at lower temperatures.

\textit{Our analytical model of the turbulent front} is based on
 the quasi-classical "coarse-grained"   equation averaged over the
 vortex lines ~\cite{Sonin,dyn-review},
   \BE{NS}
   \dot {\B U }+(1-\a')(\B U \cdot \B \nabla) \B U+ \B \nabla
  \mu =  -\G  \B U\,, \ \G\= \a \, \o\sb{ef}\,,   \ee
 which describes the evolution of  the superfluid velocity  ${\bf
U}(\B r,t) $ at scales exceeding the crossover scale  $\ell$
between hydrodynamic and kinetic regimes of turbulent motions,
which is of the order of the intervortex separation. In \Eq{NS}
$\mu$ is the chemical potential and the dissipative term $\G$ is
taken in the simplified form~\cite{LNV}, in which $\o\sb{ef}$ can
be understood as an effective vorticity.

 To estimate $V\sb f$  in the quasi-classical regime, described by
 \Eq{NS},   we consider   the total energy dissipation in the
 front with well developed turbulence which has   two
 contributions. The first one   originates from the mutual
 friction which acts on the global scale. It can be estimated from
 \Eq{NS} as $\a\o\sb{ef}K(z,r)$, where $K(z,r)= \hf\<u^2\> $ is
 the turbulent kinetic energy per unit mass and $\B u$ is the
 turbulent velocity fluctuations (with zero mean). The second
 contribution is determined by the usual
 energy flux in classical turbulence $\ve\simeq b\, K^{3/2}(\B  r)/L(\B  r)$ at an outer
 scale of turbulence $L(\B  r)$. Clearly,  $L(\B  r)\simeq \D(r)$,
 the thickness of the turbulent front at given radius $r$ near the
 centerline of the cell, or, near the surface of the cell, as
 distance to surface, $R-r$. In the whole cell, one can use an
 interpolation formula $L^{-1}(z)=\D(r)^{-1}+ (R-r)^{-1}$. The
 natural assumption is that this energy dissipates on the way to
 small scales either due to the mutual friction at moderate
 temperatures or converges into Kelvin waves at the crossover
 scale $\ell(\B r)$. For the classical Kolmogorov-41 regime
 $b\sb{cl}\simeq 0.27$ \cite{LPR}.  Using \Eq{2}   we can present
 the overall energy budget    as:
\BE{3} V\sb f \, \O^2   R^4= 8\!\! \int_0^{R-\ell}\!  r \,  d r\, dz
\Big [\G K(z,r)+ \frac{ \~ b\,  K^{3/2}(z,r)}{L(r)}   \Big]\ . 
\ee
Here we accounted for the mutual-friction correction to the
nonlinear term in \Eq{NS} $b \Rightarrow \~b\=b (1-\a') $ and used
the axial symmetry to perform the integration over the azimuthal
angle. The region with $R-r<\ell$, where
 \Eq{NS} is not applicable, is excluded from the integration.

In the turbulent boundary layer  the   kinetic energy is
independent of the axial distance to the wall. Therefore
qualitatively we can replace $K(z,r)$ and $\o \sb{ef}(z,r)$ by
their mean values across the front, $ \overline{K}(r)$ and
$\overline{\o} \sb{ef}(r)$. Also dimensional reasoning dictates $
\Delta (r)\overline{\o} \sb{ef} (r) \simeq   a \O r$   and
$\overline {K}(r)= c (\O r)^2/2 $ with $a, c \sim  1$. Now \Eq{3}
gives:
 \BE{5} 
\C V\sb f \=  V\sb f \big /  \O R\,  \simeq (2\,
c)^{3/2}b\,(1-\a')A +  4\,c a/5\, \a \,, \ee where  $ A= 0.2 +d
\big[\ln (R / \ell)  - 137/60 +5 \ell / R +\dots\big] \approx 1.8$
for  $\O=1\,$rad/s which gives $R/\ell\approx 17$. We take $b=b\sb{cl}$ and
choose the
parameters $a=0.2$, $c=0.25$ and
$d=2$ to fit the measurement in the region (0.3\,--\,0.4)$T_{\rm c}$. With these parameters  \Eq{5} gives $\C V\sb f\approx
0.16$ in the limit $T\to 0$ (when $\a=\a'=0$) and a very weak
temperature dependence up to $T\simeq 0.45T\sb c$. However, both in the region
of lower and higher temperatures the
experiment shows deviation from this
``plateau" (Fig.~\ref{f:Vf}).

The reason for this deviation at $T>0.35 T\sb c$ is that turbulence
is not well developed near the cylinder axis where the shear of
the mean velocity, responsible for turbulence excitation,
decreases. Therefore in the intermediate temperature region only
part of the front volume is turbulent, expanding toward the axis
when the temperature decreases.  We   suggest an
interpolating formula between laminar and turbulent regimes: $ \C
V\sb f=(\C V^2\sb {f, lam}+\C V^2\sb {f, turb})^{1/2}$,  where $
\C V\sb {f, turb}$ is given by \Eq{5}.  This interpolation is
shown in \Fig{f:Vf} as a thin line for $T<0.3 T\sb c$ and as a
thick line for $T>0.3 T\sb c$. The agreement with our experimental
observation for $T> 0.3 T\sb c$ is good, but there is a clear
deviation below 0.25 $T\sb c$, where $\a \lesssim 10^{-2}$.   The
reason is that we did not account adequately for the quantum
character of turbulence, which becomes important  in our
conditions  at   $T < 0.3\,T_{\rm c}$~\cite{volovik-criterium},
which  is close to the measured transition to
the lower plateau in Fig.~\ref{f:Vf}.

The mean free path of $^3$He quasiparticles at   $T\simeq
0.3\,T_{\rm c}$ is close to $\ell$ while at $0.2\,T_{\rm c}$ it
exceeds $R$. This change from the hydrodynamic to
the ballistic regime in the normal component may influence the
mutual friction force acting on the individual vortices. We
neglect this effect because for $T< 0.3\,T_{\rm c}$  the mutual
friction is already very small and does not directly affect $V\sb
f$.

At these temperatures the energy flux toward small scales propagates up to
the quantum scale $\ell$ and vortex discreteness and quantization effects
become most important. Even though some part of the energy is lost in
intermittent vortex reconnections, the dominant part proceeds to cascade
below the scale $\ell$ by means of nonlinearly interacting Kelvin waves
\cite{VinenNiemela,Vin03-PRL,5,svistunov}.  These waves are generated by
both slow vortex filament motions and fast vortex reconnection events. The
point is that Kelvin waves are much less efficient in the down-scale energy
transfer than classical hydrodynamic turbulence which leads to the
bottleneck effect increasing the kinetic energy at crossover scale up to
$\L^{10/3}$ times with respect to its K41 value~\cite{LNR}. In our
experiments $\L\=\ln(\ell / a_0) \simeq 10$ ($a_0$ is vortex-core radius)
and the inertial interval $R/\ell$ is about one decade.  Therefore the
distortion of the energy spectrum due to the bottleneck reaches the outer
scale, which leads to an essential suppression of the energy flux at given
turbulent energy or, in other words, to a decrease in the effective
parameter $b$, which relates $\ve$ and $K$. This effect is more pronounced
at low temperatures when mutual friction is small, thus $b(T)$ should
decrease with temperature. We analyze this effect with the help of the
stationary energy balance equation for the energy spectrum $E_k$ in $ \B
k$-space
$$
\frac{d \ve(k)} {d k} = - \G(T) \, E_k\,,\ \ve(k) \=
-(1-\a')\sqrt{k^{11}E_k} \, \frac{d ( E_k / k^2 ) } { 8\, d k} ,
 $$  
in which $\G(T)$, being the  damping~\eq{NS}, and  the energy flux
over scales $\ve(k)$ are taken in the Leiht differential
approximation~\cite{Leith}. In the calculations we use $L/\ell=12$ as ratio of
the outer and
crossover scales and characterize the bottleneck with the boundary condition
$E_k /[ k^3  d( E_k/k^2)/ d k ] = -4\cdot10^5$ at the crossover scale. The resulting function $b(T)$,  shown
in the insert in Fig.~\ref{f:Vf}, decreases from its classical
value $b\sb {cl}\approx 0.27$ down to $\simeq 0.1$ for $T<0.2 T\sb
c$. Now, accounting for the temperature dependence of $b(T)$ in
\Eq{5}, we get the temperature dependence of the quantum-turbulent
front shown in Fig.~\ref{f:Vf} by the bold-solid  green line,
which is in good agreement with our experimental data. This significant
decrease of the dissipation rate in the quantum regime is a consequence of  
the relative proximity of the outer and quantum
crossover scales in our measurement.

\textit{Conclusions:} We have established that conversion
of metastable vortex-free rotating $^3$He-B to stable state occurs
via propagation of a dynamic vortex structure, a vortex front,
whose nature depends on the magnitude of mutual friction
dissipation. At temperatures below $0.45\,T_{\rm c}$ sustained
turbulence appears in the front,  profoundly affecting the vortex
dynamics. Owing to the energy transfer in the turbulent cascade,
dissipation becomes temperature and mutual friction independent in
the $T\rightarrow 0$ limit. In this regime we have observed the influence of a
quantum cascade, involving individual vortices, on the global
dissipation rate.

{\it Acknowledgements}.  This work
 is supported by ULTI-4 (RITA-CT-2003-505313), Academy of
 Finland (grants 213496, 211507, 114887), and
 the US-Israel Binational Science Foundation.

\end{document}

%% file: mydefs.tex
\def \ed {\end{document}}
\def\Fbox#1{\vskip1ex\hbox to 8.5cm{\hfil\fboxsep0.3cm\fbox{%
  \parbox{8.0cm}{#1}}\hfil}\vskip1ex\noindent}  


\newcommand{\eq}[1]{(\ref{#1})}
\newcommand{\Eq}[1]{Eq.~(\ref{#1})}
\newcommand{\Eqs}[1]{Eqs.~(\ref{#1})}
\newcommand{\Fig}[1]{Fig.~\ref{#1}}
\newcommand{\Figs}[1]{Figs.~\ref{#1}}
\newcommand{\Sec}[1]{Sec.~\ref{#1}}
\newcommand{\Secs}[1]{Secs.~\ref{#1}}
\newcommand{\Ref}[1]{Ref.~\cite{#1}}
\newcommand{\Refs}[1]{Refs.~\cite{#1}}

\def\be{\begin{equation}}\def\ee{\end{equation}}
\def\bea{\begin{eqnarray}}\def\eea{\end{eqnarray}}
\def\bse{\begin{subequations}}\def\ese{\end{subequations}}
\newcommand{\BE}[1]{\begin{equation}\label{#1}}
\newcommand{\BEA}[1]{\begin{eqnarray}\label{#1}}
\newcommand{\BSE}[1]{\begin{subequations}\label{#1}}

\let \nn  \nonumber  \newcommand{\br}{\\ \nn}
\newcommand{\BR}[1]{\\ \label{#1}}
\def\hf{\frac{1}{2}}
\let \= \equiv \let\*\cdot \let\~\widetilde \let\^\widehat \let\-\overline
\let\p\partial \def\pp {\perp} \def\pl {\parallel}
\def\ort#1{\^{\bf{#1}}}
\def\Trans{^{\scriptscriptstyle{\rm T}}}
\def\x{\ort x} \def\y{\ort y} \def\z{\ort z}
 \def\bn{\bm\nabla} \def\1{\bm1} \def\Tr{{\rm Tr}}
\def\Re{\mbox{  Re}}
\def\<{\left\langle}    \def\>{\right\rangle}
\def\({\left(}          \def\){\right)}
 \def \[ {\left [} \def \] {\right ]}

\renewcommand{\a}{\alpha}\renewcommand{\b}{\beta}\newcommand{\g}{\gamma}
\newcommand{\G} {\Gamma}\renewcommand{\d}{\delta}
\newcommand{\D}{\Delta}\newcommand{\e}{\epsilon}\newcommand{\ve}{\varepsilon}
\newcommand{\E}{\Epsilon}\renewcommand{\o}{\omega} \renewcommand{\O}{\Omega}
\renewcommand{\L}{\Lambda}\renewcommand{\l}{\lambda}
\renewcommand{\t}{\tau}
\def\r{\rho}\def\k{\kappa}
\def\t{\theta } \def\T{\Theta } \def\s{\sigma} \def\S{\Sigma}

\newcommand{\B}[1]{{\bm{#1}}}
\newcommand{\C}[1]{{\mathcal{#1}}}    
\newcommand{\BC}[1]{\bm{\mathcal{#1}}}
\newcommand{\F}[1]{{\mathfrak{#1}}}
\newcommand{\BF}[1]{{\bm{\F {#1}}}}

\renewcommand{\sb}[1]{_{\text {#1}}}  
\renewcommand{\sp}[1]{^{\text {#1}}}  
\newcommand{\Sp}[1]{^{^{\text {#1}}}} 
\def\Sb#1{_{\scriptscriptstyle\rm{#1}}}